\documentclass[sigconf,authordraft]{acmart}

\AtBeginDocument{%
  \providecommand\BibTeX{
    {%
    \normalfont B\kern-0.5em{\scshape i\kern-0.25em b}\kern-0.8em\TeX
    }
  }
}

\copyrightyear{2020}
\acmYear{2020}
\setcopyright{acmcopyright}\acmConference[ICAIF '20]{ACM International Conference on AI in Finance}{October 15--16, 2020}{New York, NY, USA}
\acmBooktitle{ACM International Conference on AI in Finance (ICAIF '20), October 15--16, 2020, New York, NY, USA}
\acmPrice{15.00}
\acmDOI{10.1145/3383455.3422542}
\acmISBN{978-1-4503-7584-9/20/10}

\usepackage{multirow}
\usepackage{array}
\usepackage{booktabs}
\usepackage{colortbl}
\usepackage{arydshln}
\usepackage{paralist}
\begin{document}

\title{Predicting the Behavior of Dealers in Over-The-Counter Corporate Bond Markets}


\author{Yusen Lin}
\email{yusenlin@umd.edu}
\orcid{}
\author{Jinming Xue}
\email{jinming.xue@rhsmith.umd.edu}
\author{Louiqa Raschid}
\email{louiqa@umiacs.umd.edu}
\orcid{}
\affiliation{%
\institution{University of Maryland}
\city{College Park}
\state{Maryland}
\postcode{20742}
}

\renewcommand{\shortauthors}{Lin, et al.}

\maketitle

\section{Introduction}

Over-the-counter (OTC) refers to the process of trading (buying and selling)
securities that are not listed on a public exchange such as the New York Stock Exchange.
Understanding the trading activities of OTC dealers is crucial for market participants,
and for regulators, to better understand and monitor this largely opaque and complex market.
Our dataset is the OTC market in US corporate bonds.
The large number of bonds, low volume, and the lack of transparency and information exchange in OTC markets increase the role and importance of the dealers.

\begin{figure}[h]
\centering
\includegraphics[width=1.0\linewidth]{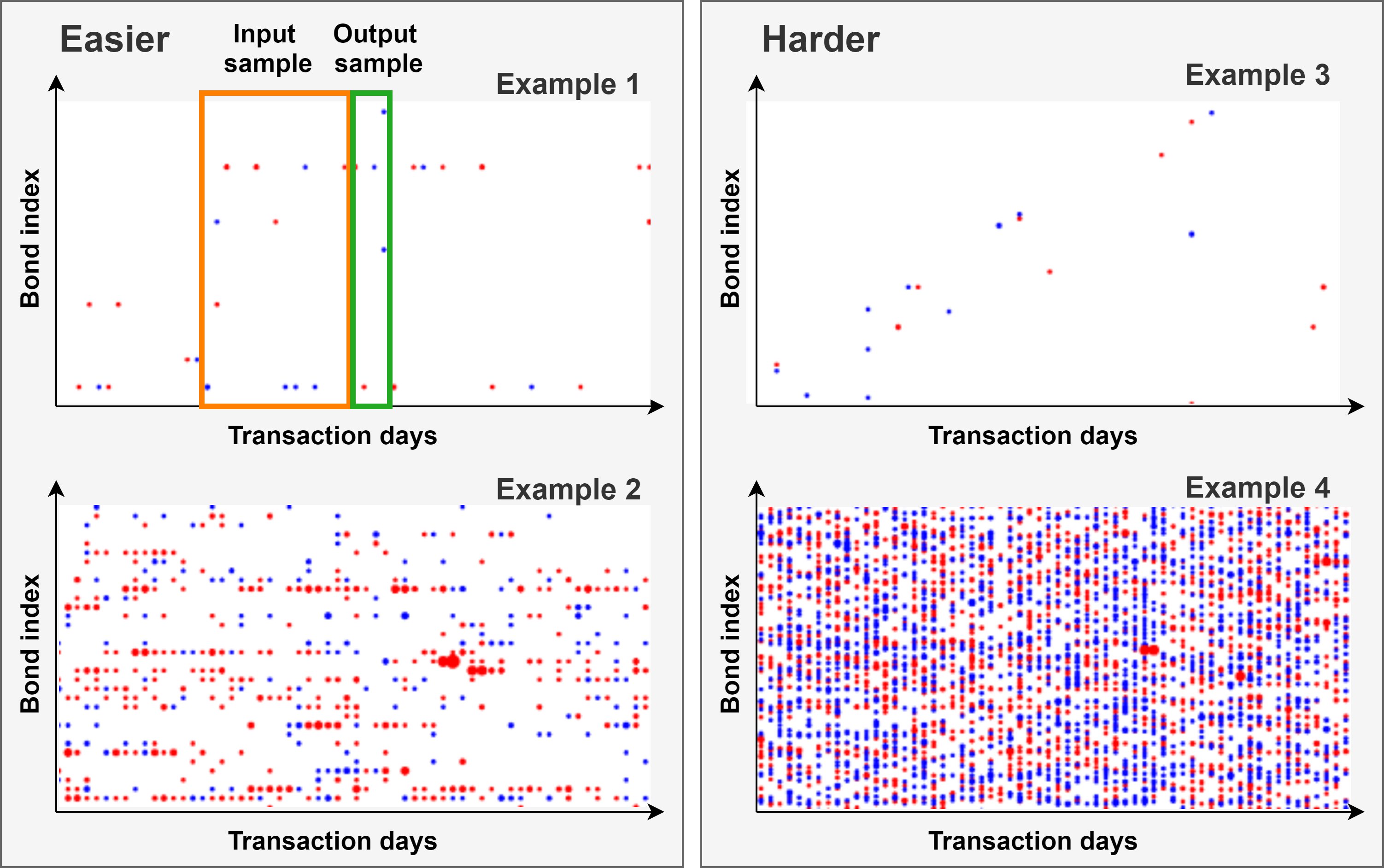}
\caption{Illustration of four dealer's trading history in 2015. 
}
\label{fig:task_description}
\end{figure}

While there has been significant research around corporate bonds, 
most of the research has used econometric models.
There is limited research on studying dealer behavior using machine learning methods.
Our objective is to predict the future behavior of individual dealers with respect to a bond.
We create the trading history of each dealer, and we use that history to create input/output training data samples.
Each input training sample represents the trades of a dealer within an input interval 
and the output sample represents their trades during the subsequent output interval. 
Then, given the input interval of a test sample, 
the model will predict the dealer's trading behavior in the output interval of that test sample.

We use the example trading activity for four dealers in Figure ~\ref{fig:task_description}
to illustrate when prediction may be accurate, and when prediction may be difficult.
The Y-axis represents the index of the bond vocabulary and the X-axis represents time.
A blue dot represents a buy action while a red one represents a sell; each horizontal line of red and blue
dots represents a dealer's history with respect to a bond.
The orange frame represents the input and the green frame represents the corresponding output, for a training or test data sample.
Dealers in Examples 1 and 2 may be easier to predict. The repeated dots across horizontal lines indicate 
repeated buy and sell actions over time for a bond; such patterns may be learned.
Examples 3 and 4 may be harder to predict. Example 3 includes only a few trades and the dealer does not seem
to repeat trades for a bond along each horizontal line. 
The history of Example 4 is dense and it may be difficult to extract relevant patterns.

\section{Dataset and Representation}
\label{sec:dataset}


Our dataset represents the OTC market in US corporate bonds.
FINRA TRACE provides granular data at the level of the individual dealer.
The TRACE academic version includes the following items: date, time, the CUSIP (identifier of the bond), and the identity of the reporting party and the counterparty.
The reporting party is a dealer while the counterparty may be a client or a dealer.  
Dealer identities are provided as an anonymized value. 
The identity of the client is not revealed.

We consider the TRACE data from 01/01/2015 to 12/31/2015.
We use a protocol from \cite{Xue2018} to filter bonds,
and the Dick-Nielsen procedure \cite{D-N2014} to delete cancellations and corrections.
There are 2.8M inter-dealer transactions, 1.7M dealer-buy-from-client transactions and 2.3M dealer-sell-to-client transactions.
The distinct count of dealers is 1.2K and the count of distinct bonds is 12K. 


We observe that the data is unbalanced. There are many dealers with low activity and a few very active dealers;
consequently, there is insufficient history for prediction for many dealers.  
There is also an imbalance in the trading activity over the bonds; this is not shown in the figure.
We, therefore, apply some additional filtering to create a subset of dealers and bonds that has 
sufficient history for prediction. 

Ideally, a prediction model should be personalized for each dealer, or more typically all the dealers
associated with a trading desk, so that we can better understand the diversity of dealer behavior.
Given the limited size of the time-series data set with at most 249 trading days in 2015, the model is very
likely to over-fit the training data, in particular for the less active dealers.
The other extreme is to combine all of the history over all dealers.  
An alternative approach is to create clusters of dealers so that we avoid over-fitting while attempting to build more personalized models; We use a set of features that are computed over the entire training interval to create four clusters.

\section{Prediction Models}

We first summarize the characteristics of various neural networks.
Then, inspired by the ReZero Transformer, we propose our Pointwise-Product ReZero (PPRZ) Transformer.

\subsection{Models}

\paragraph{Fully-Connected Neural Network (FC):} For simplicity, we use the three-layers FC networks as our baseline.

There are two variations for the input $x$ of FC.
One is $FC_{sum}$, which sums over the time-series input data, the other is $FC_{concat}$, which concatenates the input sequence.

\paragraph{RNN Based Models:} 
An RNN based model is a time-series non-linear function that recursively calculates a sequence of hidden states by converting a sequence of vectors.
We apply a Long Short-Term Memory (LSTM) based model \cite{gers1999learning} and a Bidirectional LSTM (BiLSTM) based models \cite{graves2005framewise}.

\paragraph{Transformer Based Models:} A Transformer \cite{vaswani2017attention} is good at recognizing patterns of time-series data 
by leveraging positional encoding, Self-Attention Mechanism, and Multi-Head Attention Mechanism.
In this paper, We evaluate a Vanilla Transformer \cite{vaswani2017attention} and a ReZero Transformer \cite{bachlechner2020rezero}.
The difference between a Vanilla Transformer and a ReZero Transformer is that the Layer Normalization \cite{ba2016layer} is replaced with the ReZero.

\subsection{Modification to the ReZero Transformer}

\begin{figure}[h]
\centering
\includegraphics[width=1.0\linewidth]{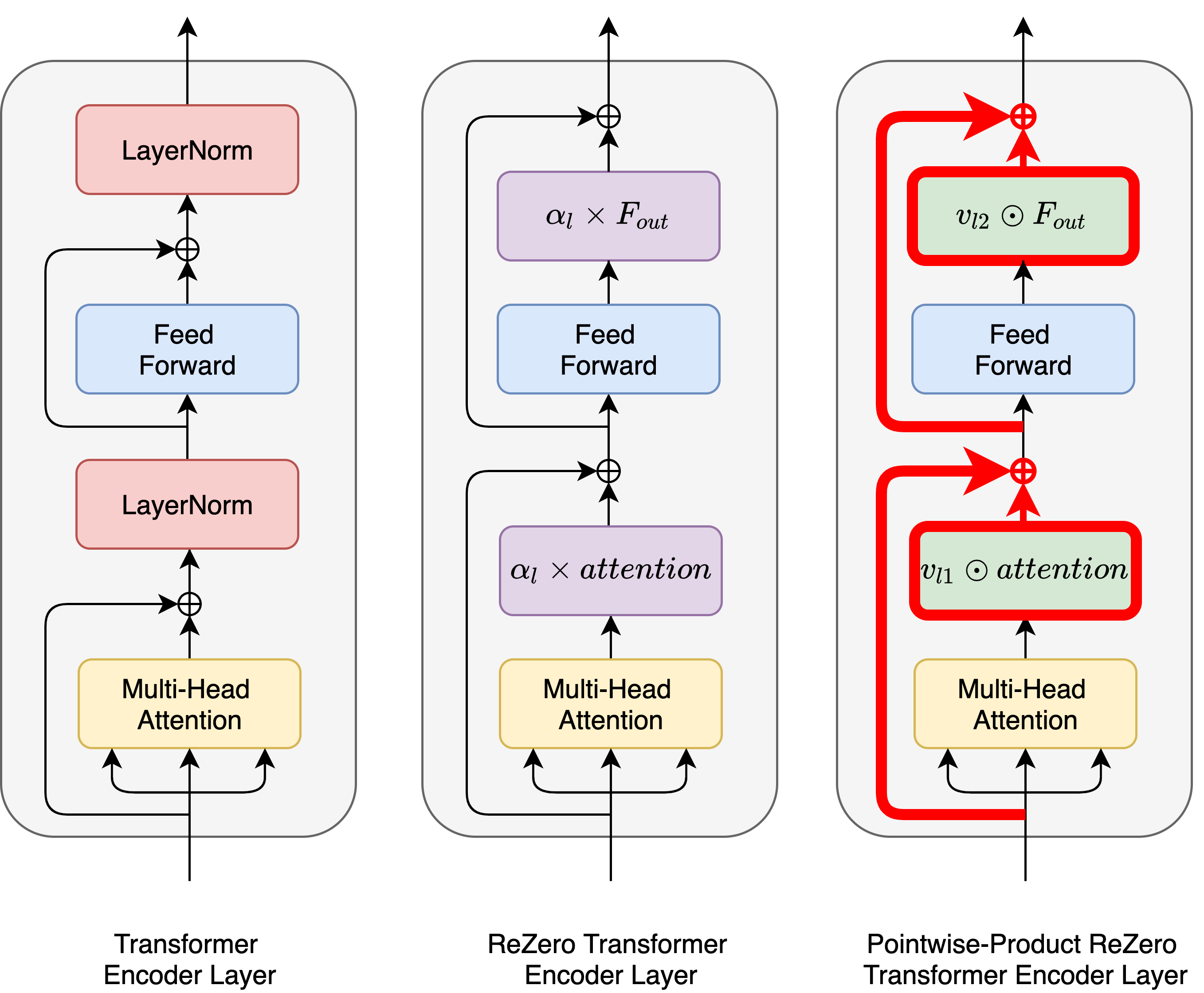}
\caption{Pointwise-Product ReZero (PPRZ) Encoder layer. We highlight the PPRZ with red color. 
Using a pointwise product with a $d$ dimension trainable vector $v_l$ can improve the normalization effect, compared to using a times operation with a scalar $\alpha_l$ in ReZero.}
\label{fig:dot_product_rezero}
\end{figure}

Inspired by \cite{bachlechner2020rezero}, we propose the Pointwise-Product ReZero (PPRZ) Transformer model.
In addition, we propose a Co-Trading Embeddings (CTE) to take advantage of distributed representations.
We use the 'tanh' function for all activation functions and apply a mean square loss to perform multi-label classification.

\paragraph{Pointwise-Product ReZero Transformer (PPRZ Transformer):} 
Figure ~\ref{fig:dot_product_rezero} illustrates the key changes to the PPRZ Transformer model. 
We replace the Layer Normalization of the Transformer.
Normalization aims to reduce 'covariate shift' \cite{ioffe2015batch} by ensuring that
signals have a zero mean and unit variance as they propagate through a network.
From the definition of the ReZero in section 3.1, we observe that by using a scalar 
multiplier $\alpha_l$, it may be difficult to reduce covariate shift.
To improve on the normalization effect, the PPRZ will
replace the scalar $\alpha_l$ and the multiplier operator.

Let $x_l$ and $F(x_l)$ be the input and the output of the $l_{th}$ layer respectively where $d$ is the last dimension.
Let $v_l$ be a $d$ dimension trainable vector $v_l$ and $\odot$ be a pointwise product.
Then PPRZ is defined as $x_{l+1} = x_l + v_l \odot F(x_l)$.
Since $v_l$ is is a trainable vector, it can be adjusted for each element of $F(x_l)$.
Our experiment results show that the output signal of the PPRZ Transformer is closer to a zero mean and unit variance, in comparison to the output signal of the ReZero Transformer.

\paragraph{Co-Trading Embedding (CTE):} 
We first create a trainable embedding of $d$ dimensions for each bond $i$ and then sum over all the traded bonds in $t_{th}$ day.
We label it as CTE since it captures features reflecting the co-trading of groups of bonds, by the same dealer.
Following the weight sharing movement in \cite{press2016using}, the CTE layer is shared between the encoder and the decoder of all Transformer based models.

\section{Experiments and Summary}

\subsection{Setup}

\noindent \textbf{Data:} As discussed earlier, we considered the 200 most active dealers, and we filtered out the most active bonds.


\noindent \textbf{Train/Test:}
We used 90\% of the data as training data, this is the interval prior to November 25.
Each training and test sample comprises an input of $T_{in}$ days and an output of $T_{out}$ days.
We vary $T_{in}$ from 5 to 10 to 15 days, to provide a diversity of training patterns.


\noindent
{\bf Models:}
We consider the following three variants for the
granularity of dealers and training data for the models:

\begin{itemize}
    \item Individual: 
    We use each individual dealer's transaction history to train a personalized model for the dealer. 
    \item Cluster:
    We combine the transaction history of all the dealers within a cluster to train the model.  
    We illustrated the range of dealer behavior in Figure ~\ref{fig:dealer_discrepancy}
    and considered properties used for clustering in Section ~\ref{subsec:clustering}. 
    \item Single:
    We train a single model on the combined transaction history of the 200 dealers.
\end{itemize}

\noindent
{\bf Metrics:}
We evaluate our models using precision, recall, and the F1 score. 


\subsection{Results}

We first compare the accuracy of prediction using the F1 score for the eight variants, and for
four clusters.
We then use the best model ($Trans_{PPRZ}$) to compare the training data variants of 
'Individual', 'Cluster', and 'Single'.

\newcommand{\tabincell}[2]{
    \renewcommand\arraystretch{1.1}
    \begin{tabular}{
    @{}#1@{}
    }#2
    \end{tabular}
}

\begin{table}[h]
  \caption{F1 Score for model variants for four clusters.
  The column index indicates the activity level of the cluster, from "least" active on the left,
  to "most" active. The box indicates the cluster with the best performance of the corresponding model; the bold highlights the best model of the corresponding cluster.
  }
  \label{tab:compare_models}
  \begin{tabular}{cccccc}
    \toprule
   F1 score & least & less & more & most & avg\\
    \midrule
    $FC_{sum}$ & 42.8 & 43.8 & 41.6 & \framebox{52.6} & 44.1 \\
    $FC_{concat}$ & \framebox{81.7} & 72.6 & 57.8 & 54.2 & 68.3 \\
    $LSTM$ & \framebox{88.8} & 76.2 & 64.9 & 56.9 & 73.7 \\
    $BiLSTM$ & \framebox{87.8} & 81.3 & 72.2 & \textbf{61.2} & 77.7 \\
    $Trans_{fv}$ & \framebox{91.7} & 81.7 & 68.8 & 61.2 & 77.9 \\
    $Trans_{CTE}$ & \framebox{91.7} & 83.6 & 70.7 & 58.8 & 78.7 \\
    $Trans_{re}$ & \framebox{86.5} & 84.1 & 74.9 & 55.1 & 78.3 \\
    $Trans_{PPRZ}$ & \framebox{\textbf{93.0}} & \textbf{88.3} & \textbf{76.1} & 60.1 & \textbf{82.3} \\
    \bottomrule
  \end{tabular}
\end{table}

\paragraph{F1 Score for Model Variants:} 
Table ~\ref{tab:compare_models} reports on the F1 score for four clusters.
The $FC_{sum}$ model exhibits the least accuracy of prediction across all four clusters.
This is not surprising since $FC_{sum}$ summarizes the time-series training input $X^t$, and is 
unable to benefit from the time-series.
This is reflected in the improved accuracy of $FC_{concat}$ over $FC_{sum}$; it utilizes a concatenation of the 
time-series training input but it cannot benefit from the ordering.
The other six models are able to utilize the full information of the time-series of the training data input and show significant performance improvement.
This confirms the importance of the time dimension on the models.

The models based on a Transformer typically achieve a higher F1 score, in comparison to  LSTM and BiLSTM.
We believe that the attention mechanism in the Transformer-based models is better able to exploit temporal
patterns as well as the importance of more recent information, in the models.
Further, the Transformer with the $CTE$ layer, $Trans_{CTE}$ performs better than the one without the $CTE$ layer, $Trans_{fv}$ .
We speculate that the $CTE$ layer can capture co-trading relationships and reduce the impact of sparse input, in the less active clusters,
leading to an improvement for these clusters.

Finally, our proposed PPRZ Transformer $Trans_{PPRZ}$ is able to outperform all of the other models, for all of the clusters, except the cluster with the most active dealers.
We note that for this most active cluster, the most accurate performance is from the BiLSTM model.
This suggests that while the $CTE$ embedding provides an advantage for the less active clusters,
this advantage is not observed in the most active clusters with sufficient training data.


\paragraph{Comparison of 'Individual', 'Cluster', and 'Single' Training Variants:}
From our preliminary experiments which report on the results of our PPRZ Transformer,
for the most active cluster, the 'Individual' variant has the best performance.
One reason is that the similarity among those most active dealers is low so that the clustering for them is ineffective for prediction.
Those most active dealers trade numerous of bonds in a day.
In addition to popular bonds, they trade bonds in various domains, and they share fewer commonalities in trading those bonds with other dealers.
For the other three clusters, the 'Cluster' models have the best performance, followed by the 'Single' variant;
the 'Individual' variant has the worst performance.
For the least active dealers, those dealers may share more commonalities with each other, or the bonds they trade are more susceptible to the same issues.
Besides, dealers from the same cluster share the same vocabulary, thus different dealers could adjust the vocabulary and make the model more robust.
Overall, the 'Cluster' level models are more effective for predicting the trading actions than the 'Individual' and 'Single' level models.

\subsection{Conclusion}

We compare the performance of a range of deep learning models 
and we propose an extension, Pointwise-Product ReZero (PPRZ) Transformer.
We demonstrate that the PPRZ Transformer has improved prediction accuracy.
There is significant variance in the level of activity of dealers; we thus considered three variants of
'Individual', 'Cluster', and 'Single', to group the training data of the dealers.


\begin{acks}
We acknowledge the support of NSF Grant OIA 1937153 and JP Morgan AI Faculty Research Award 19051422.
\end{acks}

\bibliographystyle{ACM-Reference-Format}
\bibliography{ICAIF_2020}


\end{document}